\documentstyle[psfig,epsf,axodraw]{article}
\begin{document}

\newcommand{\sla}{\kern -5.4pt /}
\newcommand{\slalarge}{\kern -10 pt /}
\newcommand{\Dir}{\kern -6.4pt\Big{/}}
\newcommand{\Dirin}{\kern -10.4pt\Big{/}\kern 4.4pt}
\newcommand{\DDir}{\kern -7.6pt\Big{/}}
\newcommand{\DGir}{\kern -6.0pt\Big{/}}

\newcommand{\ra}{\rightarrow}
\newcommand{\be}{\begin{equation}}
\newcommand{\ee}{\end{equation}}
\newcommand{\bea}{\begin{eqnarray}}
\newcommand{\eea}{\end{eqnarray}}
\newcommand{\beanon}{\begin{eqnarray*}}
\newcommand{\eeanon}{\end{eqnarray*}}
\newcommand{\ba}{\begin{array}}
\newcommand{\ea}{\end{array}}
\newcommand{\bd}{\begin{description}}
\newcommand{\ed}{\end{description}}
\newcommand{\bt}{\begin{tabular}}
\newcommand{\et}{\end{tabular}}
\newcommand{\bi}{\begin{itemize}}
\newcommand{\ei}{\end{itemize}}
\newcommand{\ben}{\begin{enumerate}}
\newcommand{\een}{\end{enumerate}}
\newcommand{\bc}{\begin{center}}
\newcommand{\ec}{\end{center}}
\newcommand{\bflr}{\begin{flushright}}
\newcommand{\eflr}{\end{flushright}}
\newcommand{\bfll}{\begin{flushleft}}
\newcommand{\efll}{\end{flushleft}}
\newcommand{\ul}{\underline}
\newcommand{\ol}{\overline}
\newcommand{\dotp}{\!\cdot\!}

\parindent 0cm
\newcommand{\parno}{\par\noindent}

\newcommand{\ar}{\rightarrow}
\newcommand{\vsk}{\vskip 10 pt\noindent}
\newcommand{\hsk}{\hskip 10 pt\noindent}
\newcommand{\vskk}{\vskip .5cm\noindent}
\newcommand{\hskk}{\hskip .5cm\noindent}

\newcommand{\ph}{{\tt PHACT\ }}
\newcommand{\phc}{{\tt PHACT}}
\newcommand{\wph}{{\tt WPHACT\ }}
\newcommand{\sph}{{\tt SIXPHACT\ }}
\newcommand{\dotab}{{\tt dotab\ }}

\def\epem{\ifmmode{e^+ e^-} \else{$e^+ e^-$} \fi}
\newcommand{\eeww}{$e^+e^-\rightarrow W^+ W^-$}
\newcommand{\qqQQ}{$q_1\bar q_2 Q_3\bar Q_4$}
\newcommand{\eeqqQQ}{$e^+e^-\rightarrow q_1\bar q_2 Q_3\bar Q_4$}
\newcommand{\eewwqqqq}{$e^+e^-\rightarrow W^+ W^-\ar q\bar q Q\bar Q$}
\newcommand{\eeqqgg}{$e^+e^-\rightarrow q\bar q gg$}
\newcommand{\eeqqqq}{$e^+e^-\rightarrow q\bar q Q\bar Q$}
\newcommand{\eewwjjjj}{$e^+e^-\rightarrow W^+ W^-\rightarrow 4~{\rm{jet}}$}
\newcommand{\eeqqggjjjj}{$e^+e^-\rightarrow q\bar 
q gg\rightarrow 4~{\rm{jet}}$}
\newcommand{\eeqqqqjjjj}{$e^+e^-\rightarrow q\bar q Q\bar Q\rightarrow
4~{\rm{jet}}$}
\newcommand{\eejjjj}{$e^+e^-\rightarrow 4~{\rm{jet}}$}
\newcommand{\jjjj}{$4~{\rm{jet}}$}
\newcommand{\qqbar}{$q\bar q$}
\newcommand{\ww}{$W^+W^-$}
\newcommand{\sm}{${\cal {SM}}$}
\newcommand{\wwqqqq}{$W^+ W^-\ar q\bar q Q\bar Q$}
\newcommand{\qqgg}{$q\bar q gg$}
\newcommand{\qqqq}{$q\bar q Q\bar Q$}

\def\Ord{\buildrel{\scriptscriptstyle <}\over{\scriptscriptstyle\sim}}
\def\OOrd{\buildrel{\scriptscriptstyle >}\over{\scriptscriptstyle\sim}}
\def\pl #1 #2 #3 {{\it Phys.~Lett.} {\bf#1} (#2) #3}
\def\np #1 #2 #3 {{\it Nucl.~Phys.} {\bf#1} (#2) #3}
\def\zp #1 #2 #3 {{\it Z.~Phys.} {\bf#1} (#2) #3}
\def\pr #1 #2 #3 {{\it Phys.~Rev.} {\bf#1} (#2) #3}
\def\prep #1 #2 #3 {{\it Phys.~Rep.} {\bf#1} (#2) #3}
\def\prl #1 #2 #3 {{\it Phys.~Rev.~Lett.} {\bf#1} (#2) #3}
\def\intj #1 #2 #3 {{\it Int. J. Mod. Phys.} {\bf#1} (#2) #3}
\def\mpl #1 #2 #3 {{\it Mod.~Phys.~Lett.} {\bf#1} (#2) #3}
\def\rmp #1 #2 #3 {{\it Rev. Mod. Phys.} {\bf#1} (#2) #3}
\def\cpc #1 #2 #3 {{\it Comp. Phys. Commun.} {\bf#1} (#2) #3}
\def\xx #1 #2 #3 {{\bf#1}, (#2) #3}
\def\prepr{{\it preprint}}
\def\hbeta{{{\beta} \over 2}}

\begin{titlepage}
\setcounter{page}{0}
\topmargin 3cm
\begin{flushright}
{\large DFTT 61/99}\\
{\rm November 1999\hspace*{.5 truecm}}\\
\end{flushright}

\vspace*{3cm}
\begin{center}
{\Large \bf \noindent  PHACT : Helicity amplitudes for present and future 
colliders
\footnote{ Work supported in part by Ministero
dell' Universit\`a e della Ricerca Scientifica\\ 
e-mail: ballestrero@to.infn.it \hskip .2cm alessandro.ballestrero@cern.ch}}
\\[1.5cm]

{\large  Alessandro Ballestrero 
   }\\[.3 cm]

{\it I.N.F.N., Sezione di Torino and \\
 Dipartimento di Fisica Teorica, Universit\`a di Torino}\\
{\it v. Giuria 1, 10125 Torino, Italy.}

\vspace*{4cm}

\centerline{\bf ABSTRACT}
\vsk
{\normalsize\noindent 
Helicity amplitudes calculations with the program \ph are explained.\\
Some examples of  
their application in   \wph and \sph MC's are given.  
} 

\vskip 5cm
{Invited talk given at QFTHEP'99 Workshop\\ 
Moscow, Russia, May 27 - June 2  1999}\\
\smallskip
{\em to be published in the proceedings}\\

\end{center}
\end{titlepage}

\begin{center}
{\Large \bf PHACT : Helicity amplitudes for present and future colliders } 

\vspace{4mm}

Alessandro Ballestrero\\
INFN and Dip. Fisica Teorica\\
via Giuria 1 - 10125 Torino\\
 Italy\\
\end{center}

\begin{abstract}
Helicity amplitudes calculations with the program \ph are explained.
Some examples of  
their application in   \wph and \sph MC's are given.  
\end{abstract}




\section{ Introduction }

The possibility of exploring higher and higher energies
at present and future accelerators, entails the necessity of predicting 
and computing with high precision more and more 
complicated processes, with
more external particles and more Feynman diagrams to be evaluated.
This has lead to progressively abandon  the old trace methods for 
perturbative computations and to use instead the helicity amplitudes ones.
\cite {gen}

 In general these can be divided in three categories, 
depending on how the various fermionic lines appearing
in a diagram are computed: by 
Trace evaluation, Spinor techniques 
or  Matrix multiplication.
We will not review here explicitly the implementations of the three methods,
but we refer to  \cite{tra}, \cite{bdks}, \cite{matr} respectively
as typical examples of them.

The program \ph \cite{ph} implements an helicity amplitude method\cite{meth} 
which belongs to both the second and third type:  one makes use
 of  a generalization of  BDKS spinors\cite{bdks}, which induce naturally
the evaluation by   multiplication of so called $\tau$ matrices. One of
the advantages of the method is that in such
a way the matrices for spinor propagators become trivial.

We will review some of its characteristics, give a short account
of \ph  and describe some of its applications.

\section{ Method } 
 
The method of ref.~\cite{meth} turns out to be easy to program, 
particularly fast and
very convenient for  computations involving massive fermions.

It is based on the following steps:
\bd
\item{-}  Generalize spinors so that $u(p,\lambda)$ and $\bar{u}(p,\lambda)$
are defined for any timelike, lightlike and spacelike vector $p$ and in all 
cases the completeness relation holds: 
\be\label{completeness}
\sum_\lambda\frac{u(p,\lambda)\bar{u}(p,\lambda)-v(p,\lambda)
\bar{v}(p,\lambda)}{2m}=1 
\ee
\item{-} Make use of the above completeness relation to diagonalize  
$p\sla$ in fermion propagators 
\item{-} Use explicitly  generalized BDKS spinors for which the following
relations hold for all   $p^2$ ($m=\sqrt{p^2}$). 
\be \hskip -2cm
u(p,\lambda)=\frac{p\sla + m}{\sqrt{2\,p\dotp k_0}}\;w(k_0,-\lambda)
\hskip 1. truecm
v(p,\lambda)=\frac{p\sla - m}{\sqrt{2\,p \dotp k_0}}\;w(k_0,-\lambda)
\ee
\be\hskip -2cm
\bar u(p,\lambda)=\bar w(k_0,-\lambda)\;\frac{p\sla + m}{\sqrt{2\,p \dotp k_0}}
\hskip 1. truecm
\bar v(p,\lambda)=\bar w(k_0,-\lambda)\;\frac{p\sla - m}{\sqrt{2\,p \dotp k_0}}
\ee
with
\be\hskip -1cm
w(k_0,\lambda)\bar{w}(k_0,\lambda)=\frac{1+\lambda\gamma_5}{2}k\sla_0
\hskip 1cm
w(k_0,\lambda)=\lambda k\sla_1 w(k_0,-\lambda)
\ee
\[ k_0^2=0 \hskip 1.5cm k_1^2=-1 \hskip 1.5cm k_1\dotp k_0=0\]
\ed

As a consequence, we have that
\bd
\item{-} Every insertion of a vector (or scalar) $\eta_\mu$  in a fermion line

\begin {center}
\begin{picture}(80,35)(0,0)
\SetWidth{1.2}
\SetOffset(0,0)
\Line(0,0)(60,0)
\Text(-6,-8)[]{$M_1$}
\Text(66,-8)[]{$M_2$}
\Text(-6,+8)[]{$p_1$}
\Text(66,+8)[]{$p_2$}
\Text(-60,0)[]{$M_1=\sqrt{p_1^2}$}
\Text(120,0)[]{$M_2=\sqrt{p_2^2}$}
\Photon(30,0)(30,40){2}{3}
\Text(43,40)[]{\large $\eta_\mu$}
\end {picture}
\end{center}

\be 
\ol {U}(p_1)\eta\sla \left[
c_{r} \left(\frac{1+\gamma_5}{2}\right) +
             c_{l} \left(\frac{1-\gamma_5}{2}\right) \right]U(p_2)
   \ee
turns out to be of the form (also for scalar coupling):
\be A+B\, M_1+C \,M_2 + D\, M_1 M_2 \ee
with {\sl A, B, C, D} \hsk 2x2 matrices (+ and - correspond to the values
of the helicity $\lambda$ indices) :
\be \left(\ba{cc}A_{++}&A_{+-}\\A_{-+}&A_{++}\ea\right) \ee 

Here $\eta_\mu$ represents a polarization vector or a whole subdiagram with 
appropriate indices.

\item{-} Every fermion line or piece of it

\begin {center}
\begin{picture}(500,40)(0,0)
\SetWidth{1.2}
\SetOffset(0,+6)
\ArrowLine(100,0)(0,0)
\Photon(20,0)(20,40){2}{3}
\Photon(50,0)(50,40){2}{3}
\Photon(80,0)(80,40){2}{3}
\Text(-6,-8)[]{$M_1$}
\Text(106,-8)[]{$M_2$}

\SetOffset(180,6)
\ArrowLine(60,0)(0,0)
\Photon(20,0)(20,40){2}{3}
\Photon(40,0)(40,40){2}{3}

\Text(65,0)[lc]{....}

\SetOffset(270,6)
\ArrowLine(60,0)(0,0)
\Photon(20,0)(20,40){2}{3}
\Photon(40,0)(40,40){2}{3}
\Text(-6,-8)[]{$M_1$}
\Text(66,-8)[]{$M_2$}

\Text(70,0)[lc]{....}

\SetOffset(355,6)
\ArrowLine(60,0)(0,0)
\Photon(20,0)(20,40){2}{3}
\Photon(40,0)(40,40){2}{3}
\end {picture}
\end{center}

is of the same form:
\be A+B\, M_1+C \,M_2 + D\, M_1 M_2 \ee
It can be represented by a $\tau$ matrix: 
\be\tau=\left(\ba{cc}A&C\\B&D\ea\right)\ee

\item{-} Composition of two pieces of fermion line is trivial:

\begin {center}
\begin{picture}(500,60)(0,0)
\SetWidth{1.2}
\SetOffset(105,20)
\Text(0,0)[]{{\large$\tau =$}}
\ArrowLine(20,0)(200,0)
\Photon(50,0)(50,40){2}{3}
\Text(60,-15)[]{\large{$\tau_1$}}
\Photon(70,0)(70,40){2}{3}
\Text(110,10)[]{\large{$p\sla + \mu$}}
\Photon(150,0)(150,40){2}{3}
\Photon(170,0)(170,40){2}{3}
\Text(170,-15)[]{\large{$\tau_2$}}
\Photon(190,0)(190,40){2}{3}
\end {picture}
\end{center}
\large
\be\tau=\tau_1\otimes\tau_2\]
\normalsize
\vsk
\[
\left(\ba{cc}A&C\\B&D\ea\right)  =
\left(\ba{cc}A_1&C_1\\B_1&D_1\ea\right)  
\left(\ba{cc}1&\mu\\\mu&p^2\ea\right)
\left(\ba{cc}A_2&C_2\\B_2&D_2\ea\right) 
\ee
\ed
With the above results and the expressions of {\sl A, B, C, D}\cite{meth}
 for a single vector or scalar insertion, any calculation can in principle be
performed.
This  way of evaluating amplitudes has some advantages as it 
 eliminates the combinatorics induced by expressing propagator momenta
in terms of other 4-vectors and it allows a natural decomposition in 
subdiagrams. This in turn 
 reduces the number of graphs to account for and easily
 avoids repeating computations of subgraphs common to different diagrams.

A typical example of modular computation with \ph is described in 
\cite{sixphact}: for the process 
 $\epem \ar \mu \nu u \bar d b \bar b$ 
which has 8 external fermions and 209 Feynman diagrams 
we have computed the amplitude by first evaluating the  subdiagrams
corresponding to the emission of 4 fermions from a virtual vector boson and to
the emission of four fermion from a fermion line. With these subdiagrams
we end up with only 18 diagrams.

\section{\ph } 

  \ph ({\bf P}rogram for {\bf H}elicity {\bf A}mplitude {\bf C}alculations with
{\bf T}au matrices)\cite{ph}  is a program to write optimized fortran 
codes for helicity amplitude processes with the above described method.
It is composed of many routines which write the code for the different
parts: $\tau$ matrices for a single insertion, products $\tau \otimes \tau$
of matrices, vector boson polarizations, 
  triple vertices, determinants, etc. 
In all cases one gives as arguments
to the routines the names  to be used for vectors, matrices, couplings, etc.
and the code will be written with the appropriate variables. 

The purpose is to generate  a final code in the easiest way.
Instead of generating all the fortran program with \phc, one could also 
think of 
another strategy. 
At least in some simple cases one could in fact write a fortran program 
where one subroutine (say for instance the one to compute $\tau$) is called 
many times.   However just in the case at hand  
 the ``polarization'' vector $\eta_\mu$ can correspond  to different 
subdiagrams, and as such it can carry different number of indices. This is
not naturally and easily implemented in a repeated call to a fortran routine,
while with \ph one can produce the code for each different call with its
own indices and the resulting program will be much less time consuming.

Moreover, writing a code with the help of \ph is greatly simplified 
by some useful  features:
\bd
\item{-}    normal indented fortran lines can be alternated to specific calls
   to \ph in the input program, and they will be recognized as such
   and left unchanged.  So after the input has been processed by \phc,
   the resulting  final fortran code will be composed by the parts written
   directly in the input and by those  generated automatically.     
\item{-} when one wants to repeat a piece of input just changing some names
       or indices, one can use a \dotab command.
      After the command, in the first line are given the symbols to
      change and in the following ones the values they will assume 
      at the first, second, ... repetition. {\tt endtab} terminates the list
      and {\tt enddo} terminates the part of input code to be repeated. 
      {\tt dotab} substitutions can be applied also to normal fortran lines 
in  input and can be nested.
\item{-} for many \ph routines the use of indices followed by a question mark ? 
       will result in fortran do loops on those indices in the
      code generated by the routine. 

\ed

The  subroutines which can be called using \ph are at present the following:

\begin{verbatim}
pol,pk0,quqd,p.q,eps,trip,mline
t,tsts,tst,tts,tt 
t0,tl0,tr0,t10,tlt0,ttr0,tltr0
tw,tstw,twts,twt,ttw,twtw
tw0,twl0,twr0,tw10,tlt0_w,ttr0_w,tltr0_w
th,tsc,tsct,ttsc,tscts,tstsc,tsctw,twtsc,tsctsc
tot,tto,tots,tsto,totw,twto,toto,totsc,tscto
tcr,tcrtcr
\end{verbatim}

By calling the ones in the first line and specifying their arguments, 
one obtains the fortran code for computing scalar products, determinants,
polarization vectors, triple vertices etc.
In the second line the routine {\tt t} writes the fortran code for the
$\tau$ matrix of one single insertion, and the others combine
different $\tau$'s together. Similar routines in the third line
serve for the case of massless fermion lines. In the two following
lines are the corresponding routines for insertions of W's (left
coupling only), while the remaining ones are for scalars, pseudoscalars and
right couplings. 

As a simple example of the use of these routines let's consider how
one gets the code for the single insertion:

\begin{verbatim}
t        !subroutine T [qu,qd,v,abcd,cr,cl,den,nquqd,nsum]
p1
p2
v34(i,j).e
res(i?,j?).
coupr
coupl
0
0
0
\end{verbatim}

After calling the routine {\tt t}, one gives as arguments
the names of the 4-momenta of the fermion line, the name
of the ``polarization vector'' (which may correspond to a subdiagram)
 of the insertion with the corresponding indexes and the name
of the couplings. The three last entries correspond to the
possibility of dividing by a denominator, of performing
the scalar product of the two 4-vectors and of summing the
result with a previous one.

The output by \ph corresponding to the above input is:

\begin{verbatim}      
      do i=1,2                                                  
      do j=1,3                                                  
* T -- qu=p1,qd=p2,v=v34(i,j).e,a=res(i,j).a,b=res(i,j).b,c=res(i,j).c,d=re
* s(i,j).d,cr=coupr,cl=coupl,nsum=0                                         
      eps_0=-v34(i,j).ek0*(p1(2)*p2(3)-p2(2)*p1(3))+p1k0*(v34(i,
     & j).e(2)*p2(3)-p2(2)*v34(i,j).e(3))-p2k0*(v34(i,j).e(2)*p1
     & (3)-p1(2)*v34(i,j).e(3))                                 
      ceps_0=eps_0*cim                                          
      eps_1=-v34(i,j).e(3)*p1k0+p1(3)*v34(i,j).ek0              
      ceps_1=eps_1*cim                                          
      eps_2=-v34(i,j).e(3)*p2k0+p2(3)*v34(i,j).ek0              
      ceps_2=eps_2*cim                                          
      vqu=v34(i,j).e(0)*p1(0)-v34(i,j).e(1)*p1(1)-v34(i,j).e(2)*
     & p1(2)-v34(i,j).e(3)*p1(3)                                
      vqd=v34(i,j).e(0)*p2(0)-v34(i,j).e(1)*p2(1)-v34(i,j).e(2)*
     & p2(2)-v34(i,j).e(3)*p2(3)                                
      auxa=-v34(i,j).ek0*quqd+p1k0*vqd+p2k0*vqu                 
      auxb=-v34(i,j).ek0*p2(2)+p2k0*v34(i,j).e(2)               
      auxc=+v34(i,j).ek0*p1(2)-p1k0*v34(i,j).e(2)               
      res(i,j).a(1,1)=coupr*(auxa+ceps_0)                       
      res(i,j).a(2,2)=coupl*(auxa-ceps_0)                       
      res(i,j).b(1,2)=coupl*(auxb-ceps_2)                       
      res(i,j).b(2,1)=coupr*(-auxb-ceps_2)                      
      res(i,j).c(1,2)=coupr*(auxc+ceps_1)                       
      res(i,j).c(2,1)=coupl*(-auxc+ceps_1)                      
      res(i,j).d(1,1)=coupl*v34(i,j).ek0                        
      res(i,j).d(2,2)=coupr*v34(i,j).ek0                        
      end do                                                    
      end do                                                    
\end{verbatim}

Notice the use of ? which has produced the {\tt do i} and 
{\tt do j} commands. 

If  the same input lines for \ph had been preceded by
\begin{verbatim}
dotab
1  2  34   
1  2  34 
3  4  56 
5  6  78 
endtab
\end{verbatim}
and followed by
\begin{verbatim}
enddo
\end{verbatim}
in the output the fortran lines above would have been followed by
two repetitions, one in which 1 would have been changed to 3, 2 to 4
and 34 to 56, the second with 1 to 5, 2 to 6 and 34 to 78.

\section{  Applications } 
Since the first calculation performed for the process 
$e^+e^-\rightarrow b\bar bW^+W^-$\cite{prima}, 
\ph has been continuously used and tested in recent years 
in many high energies computations. In particular the two MonteCarlo's,
\wph\cite{wphact} and \sph\cite{sixphact}\cite{sixhiggs} have been written 
with it.

These MC's compute respectively 4 and 6 fermion final states and they
 have been used for phenomenological applications
and  compared with results from several other groups
in LEP2 and Linear Collider workshops. 

The method has thus proven to be able to deal with thousands of Feynman diagrams
producing  high precision results in short CPU times.

As a full description of the characteristics of these programs is far
beyond  the scope
of the present review, only some relevant and more recent features of them
 will be sketched in the rest of the section. 

\subsection{ \wph }
\wph  computes all SM 4 fermion processes including 
Higgs and MSSM Higgs diagrams.

In the first version all fermions except the b's were considered massless.
For most WW and ZZ studies this approximation is completely adequate and
allows much faster computations.

For particular processes with at least one final state electrons at low angle
or lost in the beam pipe, massless approximations are of course
not viable, as the matrix elements diverge in this approximation because of
 the
t channel photon propagator. Moreover exact gauge invariance
controls the behaviour of such a divergence and even a small
violation results in huge numerical differences.    

For such a reason and also in order to regulate other possible divergent 
behaviour at small invariant masses, completely massive matrix elements for 
all final states 
have recently been added.
Therefore, using \wph  one can now choose between
\bi
 \item massless  
 \item only b massive
 \item fully massive
\ei
matrix elements, depending on the problem at hand.

Correspondingly, phase spaces appropriate for  forward region evaluations have
been provided.

As an example  of these recent improvements, we show in fig.~1 some comparisons
at LC energies for the so called single W process
$\epem \ar e \bar\nu u \bar d $ (CC20) with the final $e$ at small angle. 
We notice that even in this difficult region and at high energy  
the agreement between the massive codes 
\wph, {\tt COMPHEP}{\cite{com} and {\tt GRC4F}\cite{grc} is of the order of a 
few per mille.

\begin{figure}[thb]
\centerline{\hskip-1cm\psfig{figure=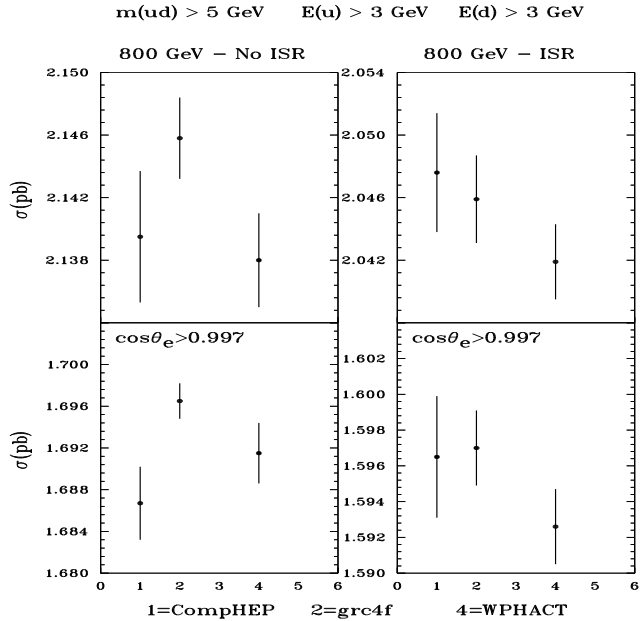,height=10cm,width=10cm}}
\vskip -1.2cm
\caption{\label{f1} 
Comparison of the predictions from {\tt COMPHEP}, {\tt GRC4F}  
and \wph for the process $\epem \ar e \bar\nu u \bar d $
with the electron in the forward region.} 
\end{figure}

\subsection{ \sph }

Six fermion final states will be relevant at LC for all studies of top, 
 intermediate Higgs, WWZ productions and anomalous couplings (quartic coupling).

As already proven by the studies at LEP2, also for these processes
the production times decay approximation will not be sufficient,
as it  ignores finite width
effects and spin correlations  and it is not suited for 
irreducible background evaluation.

\sph\ can at present compute all Charged Currents six fermion final states,
which are the states that can be produced by the decay of two W's and
a neutral particle, but not from three neutral particles. 
 These are the cleanest final states to study  intermediate Higgs, 
WWZ, $t\bar t$ physics:
 with one lepton and a neutrino (missing energy)
one automatically eliminates the most relevant part of QCD (6 jets, 4 jets +
$l\:\bar l$, ..) and ZZZ background.

Even if the number of Feynman diagrams for 6 fermion processes is one order
of magnitude greater than for 4 fermions, \sph can easily reach per mille
accuracy in a few hours at present workstations.

From a technical point of view it is interesting to point out that both
\wph and \sph can   produce automatically 
 any distribution the user defines.
Even for \sph 
high statistical accuracy  is easily achieved in these distributions.
An example of this feature is given in fig.~2, where background and
signal distributions for an intermediate higgs is studied in events
with one neutrino, one lepton and four jets. It has to be noticed
that the statistical errors on the single bins  are not visible in the figure 
as they are   below the percent.

\begin{figure}[thb]
\centerline{\hskip-.8cm\psfig{figure=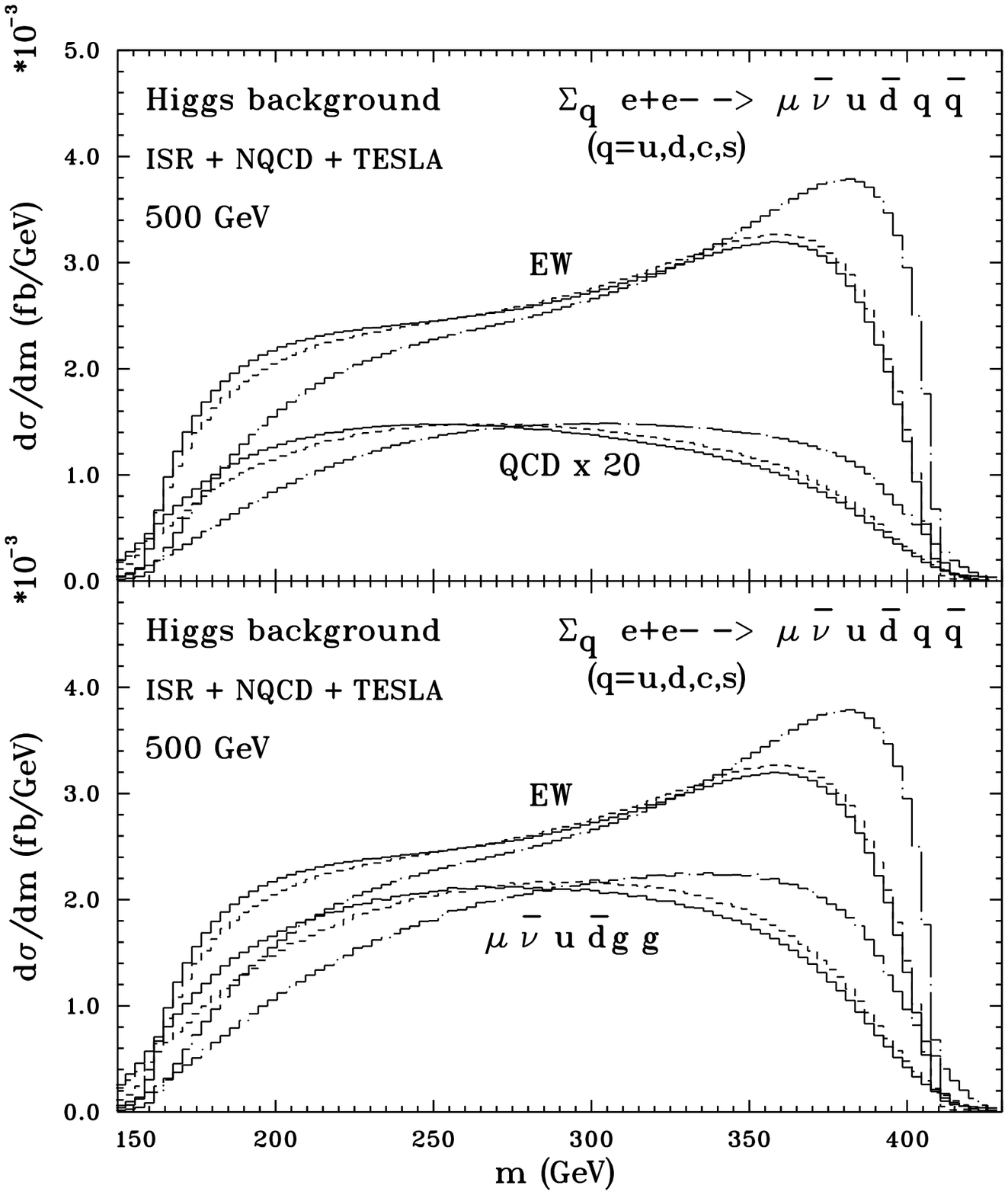,height=12cm,width=8cm}
\psfig{figure=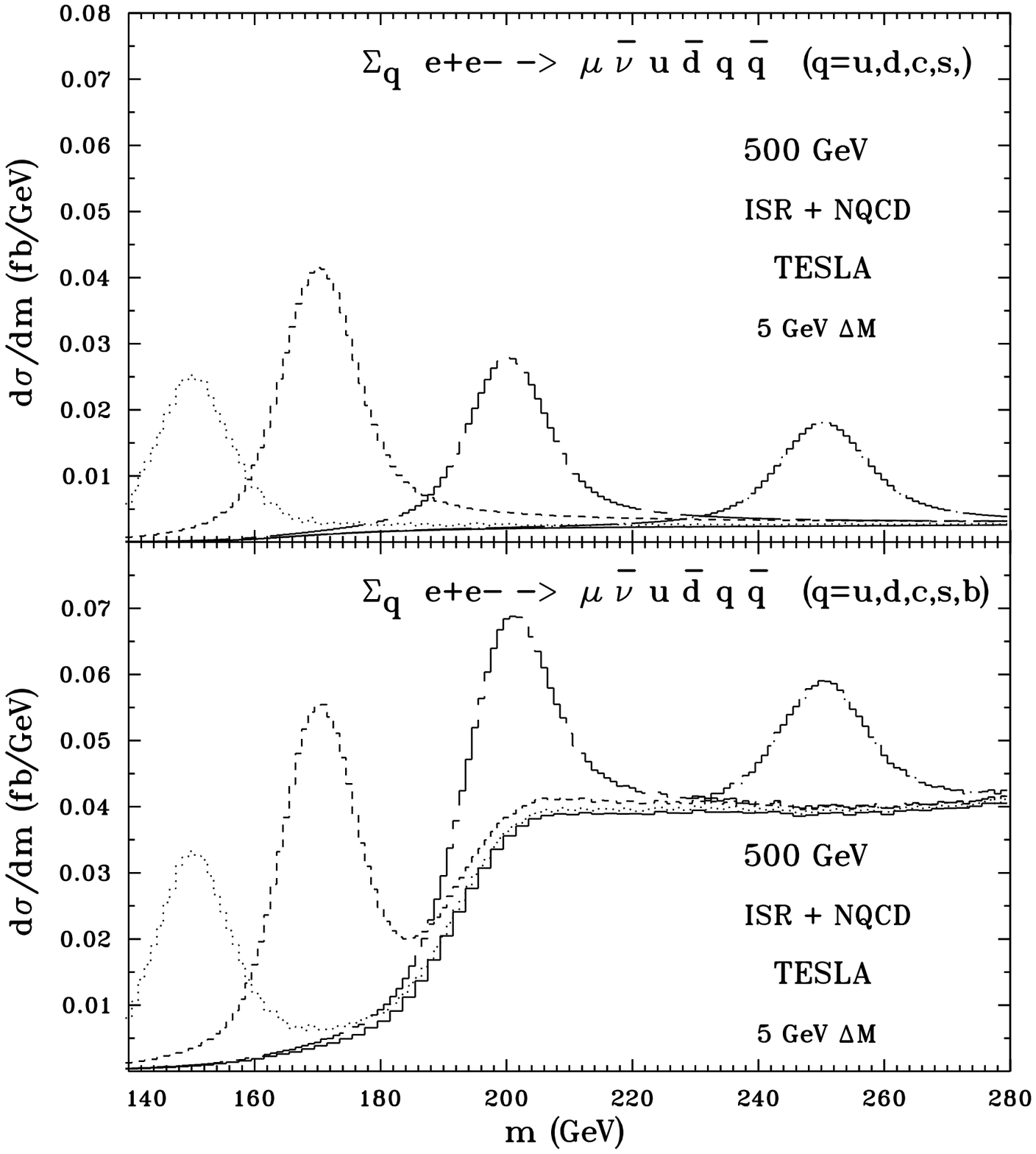,height=12cm,width=8cm}}
\vskip -.8cm
\caption{\label{f2} 
Invariant mass distributions for intermediate higgs background (left) 
and signal (right).
For the background,  continuous line  corresponds to ($\mu \bar \nu
u \bar d$)  mass, dashed  to {\it reconstructed}
and  chaindot to the {\it missing}.
On the right reconstructed mass distributions with gaussian smearing
are reported and 
the continuous line represents the total
background. The others correspond to the total cross sections for
 $m_h$= 150, 170, 200, 250 GeV.
From ref.~\cite{sixhiggs} 
}
\end{figure}

\section{Conclusions } 

We have briefly described a method for high energy computations,
the way it can be programmed and some of its realizations.

In recent years computer CPU performances have continuously improved. 
Normal workstations are now faster by about a  factor of twenty  
than  five years ago.
One may therefore wonder whether it is still important to find
faster and easier computational methods in presence of
 such technical achievements.
The answer to this question becomes evident if one considers  that 
at the same time the continuous improvements in statistics, luminosity and energy
of the experiments and of the accelerators always demand for higher
theoretical precision.  The complexity of the problems and processes
in the future is orders of magnitude greater than at present 
and therefore it will be necessary to 

\bi
\item extend automatic computations
\item find new techniques for the higher complexity
\item continue comparing results of the different approaches
\ei

\ph is intended as a tool that can contribute to this progress.
It  has so far allowed the construction
of  fast and precise MC's for \epem colliders.
In the future it will hopefully be improved and it will also be used for 
LHC physics. 

\vskip 1cm
{\bf Acknowledgment}
I am grateful to the organizers of QFTHEP'99 workshop, and in particular to
Edward Boos and Mikhail Dubinin,  for the pleasant and constructive atmosphere
during the workshop and for the warm hospitality.  


\end{document}